\documentclass[conference]{IEEEtran}
\IEEEoverridecommandlockouts
\usepackage{cite}
\usepackage{amsmath,amssymb,amsfonts}
\usepackage{algorithmic}
\usepackage{graphicx}
\usepackage{textcomp}
\usepackage{xcolor}
\def\BibTeX{{\rm B\kern-.05em{\sc i\kern-.025em b}\kern-.08em
    T\kern-.1667em\lower.7ex\hbox{E}\kern-.125emX}}

\usepackage[utf8]{inputenc}
\usepackage{physics}
\usepackage{graphicx}
\usepackage{multirow}
\usepackage{hyperref}
\hypersetup{colorlinks=true, citecolor=blue}
\usepackage{cleveref}
\usepackage{xcolor}
\usepackage{bbold} 
\usepackage{tabularx} 
\usepackage{soul}  
\usepackage{tikz}
\usetikzlibrary{quantikz2}
\usepackage{comment}
\usepackage{graphicx}
\usepackage{booktabs}
\usepackage{tcolorbox}

\usepackage[caption = false]{subfig}

\usepackage{makecell} 
\usepackage{mathtools}
\usepackage{tabularx} 
\usepackage{amsmath}
\usepackage{amssymb}
\usepackage{amsthm}
\usepackage{braket}
\usepackage{algorithm}
\usepackage{algorithmic}

\usepackage{listings}
\usepackage{caption}
\newcolumntype{C}[1]{>{\centering\arraybackslash}p{#1}}

\lstdefinelanguage{yaml}{
  keywords={true,false,null},
  keywordstyle=\color{blue}\bfseries,
  comment=[l]{\#},
  commentstyle=\color{gray}\itshape,
  string=[s]{""}{"},
  morestring=[s]{'}{'},
  sensitive=true
}

\begin{document}


\title{Kubernetes-Orchestrated Hybrid Quantum–Classical Workflows}







\author{
\IEEEauthorblockN{Mar Tejedor\IEEEauthorrefmark{1}\IEEEauthorrefmark{2}, 
Michele Grossi\IEEEauthorrefmark{3}, 
Cenk T\"uys\"uz\IEEEauthorrefmark{3}, 
Ricardo Rocha\IEEEauthorrefmark{3}, 
Sofia Vallecorsa\IEEEauthorrefmark{3}}
\IEEEauthorblockA{\IEEEauthorrefmark{1}Barcelona Supercomputing Center (BSC), Barcelona, Spain}
\IEEEauthorblockA{\IEEEauthorrefmark{2}Universitat Politècnica de Catalunya (UPC), Barcelona, Spain}
\IEEEauthorblockA{\IEEEauthorrefmark{3}European Organisation for Nuclear Research (CERN), Espl. des Particules 1211 Geneva 23, Switzerland}
}

\maketitle

\begin{abstract}
Hybrid quantum–classical workflows combine quantum processing units (QPUs) with classical hardware to address computational tasks that are challenging or infeasible for conventional systems alone. Coordinating these heterogeneous resources at scale demands robust orchestration, reproducibility, and observability. Even in the presence of fault-tolerant quantum devices, quantum computing will continue to operate within a broader hybrid ecosystem, where classical infrastructure plays a central role in task scheduling, data movement, error mitigation, and large-scale workflow coordination.

In this work, we present a cloud-native framework for managing hybrid quantum–HPC pipelines using Kubernetes, Argo Workflows, and Kueue. Our system unifies CPUs, GPUs, and QPUs under a single orchestration layer, enabling multi-stage workflows with dynamic, resource-aware scheduling. We demonstrate the framework with a proof-of-concept implementation of distributed quantum circuit cutting, showcasing execution across heterogeneous nodes and integration of classical and quantum tasks. This approach highlights the potential for scalable, reproducible, and flexible hybrid quantum–classical computing in cloud-native environments.
\end{abstract}



\section{Introduction}\label{sec:intro}

Quantum computing continues to advance as a promising computational paradigm, with constant improvements in qubit counts, connectivity, and coherence times, hybrid quantum–classical workflows, where quantum processors are tightly integrated with classical preprocessing, optimization, simulation, and post-processing~\cite{McClean_2016,CRANGANORE2024346} emerge as a persistent execution model.
Such workflows require coordinating heterogeneous resources, including CPUs, GPUs, and Quantum Processing Unit (QPUs), in a manner that is scalable, portable, and reproducible across High-Performance Computing (HPC) clusters and cloud platforms.

Cloud-native orchestration platforms provide a flexible software layer for managing heterogeneous resources. Kubernetes~\cite{burns2016design}, supported by workflow engines such as Argo Workflows~\cite{argo_workflows} and batch-scheduling layers like Kueue~\cite{kueue}, enables containerized execution, resource-aware scheduling, and unified lifecycle management for distributed scientific workloads. These characteristics align naturally with the demands of hybrid quantum computing, where tasks must span classical compute nodes, GPU-accelerated simulators, and on site or remote quantum backends provided by cloud vendors~\cite{mckay2018qiskitbackendspecificationsopenqasm,10.1145/3726301.3732296}.

In this work, we introduce a Kubernetes-native framework for orchestrating hybrid quantum--classical pipelines. The system unifies CPUs, GPUs, and QPUs within a single operational model and supports reproducible, multi-stage quantum workflows. While we illustrate the approach using a representative quantum application, the proposed architecture is general and applicable to a broad class of hybrid workloads.
This work does not aim to benchmark performance against traditional HPC schedulers, nor to propose new quantum algorithms. Instead, it focuses on orchestration, reproducibility, and workflow-level integration of heterogeneous quantum and classical resources.

\section{Related Work}

Scalable orchestration of hybrid quantum--classical workflows remains largely unexplored, particularly at the intersection of HPC, cloud-native infrastructure, and quantum middleware. Existing systems typically address isolated aspects of the problem, such as resource heterogeneity, quantum task scheduling, or general-purpose cloud orchestration, but rarely provide an integrated end-to-end workflow model.

Early work toward cloud-native quantum execution~\cite{grossi2021serverlesscloudintegrationquantum} demonstrated one of the first Kubernetes-based prototypes for deploying quantum workloads, showing the feasibility of treating quantum programs as cloud services. At the middleware level, Pilot-Quantum~\cite{mantha2024pilotquantum} offers unified scheduling across CPUs, GPUs, and QPUs, but does not consider full hybrid pipelines or cloud-native workflow integration. Classical distributed workflow systems such as ColonyOS~\cite{kaellman2025colonyos} introduce reproducibility and monitoring capabilities, yet remain limited to non-quantum workloads.

Closer to cloud-native quantum orchestration, Qubernetes~\cite{qubernetes2024} extends Kubernetes with custom resource definitions for representing quantum jobs as first-class objects. While it proves the architectural viability of managing quantum workloads in Kubernetes, it focuses on isolated quantum tasks and lacks support for multi-stage hybrid workflows, heterogeneous coordination, or scalable performance management.

No existing framework integrates hybrid quantum–classical execution, HPC resources, and cloud-native orchestration within a unified system. The framework presented in this work moves beyond single-job execution by enabling complete end-to-end hybrid pipelines, dynamic scheduling across heterogeneous resources, and integrated monitoring and reproducibility. By embedding these capabilities directly into Kubernetes, it establishes a scalable and transparent foundation for executing large-scale quantum–HPC experiments in modern distributed environments.

\section{Background}\label{sec:background}
Hybrid quantum--classical workflows combine quantum processing units (QPUs) with classical computing resources to address problems that are challenging for purely classical systems. A prominent class of such approaches is variational quantum algorithms, including the variational quantum eigensolver and the quantum approximate optimization algorithm~\cite{McClean_2016,CRANGANORE2024346}. 
In these workflows, computation is divided between classical and quantum components in an iterative feedback loop. Classical resources perform tasks such as circuit generation, compilation, parameter optimization, error mitigation, and post-processing. The QPU executes parameterized quantum circuits and returns measurement samples or expectation values, which are then used by the classical optimizer to update the circuit parameters. This hybrid structure enables the use of near-term quantum devices while leveraging mature high-performance classical infrastructure.

Beyond variational approaches, fully hybrid algorithms increasingly drive practical progress. Circuit cutting decomposes large circuits into smaller subcircuits executed independently and later recombined through classical post-processing~\cite{tejedor2025distributed,harada2023wirecut,uchehara2022ricco,bechtold2023nonmax}. Sample-based diagonalization \cite{sample-based-diagonalization} and related Hamiltonian-processing methods follow a similar pattern, interleaving quantum sampling with substantial classical computation. These and other hybrid techniques illustrate the need for coordinated execution and data flow between classical and quantum resources to achieve scalability and accuracy.

\subsection{Heterogeneous Resources}
Hybrid quantum–HPC workflows rely on a combination of classical and quantum resources, each contributing distinct capabilities to the execution pipeline. \textbf{CPUs} manage control flow, data pre- and post-processing (including error correction and mitigation), and lightweight quantum simulations~\cite{McClean_2016}. \textbf{GPUs}, with highly parallel architectures, accelerate large-scale classical simulations, tensor contractions, and statevector or density matrix propagation~\cite{10.1145/3676641.3715984}. \textbf{QPUs} execute quantum circuits on physical hardware, enabling sampling, optimization, and simulation tasks that can provide advantage under suitable conditions~\cite{Preskill2018quantumcomputingin,mckay2018qiskitbackendspecificationsopenqasm}.

Coordinating these heterogeneous resources is challenging due to differing performance characteristics, memory hierarchies, execution times and access constraints. QPU execution is typically limited by queueing delays and shot counts, while GPU simulations can be resource- and energy-intensive and also limited resources. CPUs remain essential for orchestration, data aggregation, and managing asynchronous interactions between classical and quantum stages~\cite{10.1145/3726301.3732296}.

Unified orchestration frameworks mitigate these challenges by automating resource coordination, scheduling, and monitoring across CPUs, GPUs, and QPUs~\cite{mantha2024pilotquantum}. Such frameworks maximize utilization, reduce idle times, and enable reproducible, scalable execution of hybrid workflows across heterogeneous infrastructures, providing a transparent bridge between classical and quantum computing resources.

\subsection{Kubernetes for Scientific Workflows}

Kubernetes has emerged as a leading cloud-native platform for orchestrating complex scientific workflows~\cite{burns2016design}. By abstracting infrastructure through constructs such as \texttt{Pods}, \texttt{Jobs}, and custom \texttt{Operators}, Kubernetes enables researchers to package tasks in containers, ensuring portability, reproducibility, and seamless deployment across heterogeneous environments~\cite{argo_workflows}. 

Modern scientific workflows often span diverse hardware, including CPUs, GPUs, and specialized accelerators such as AI or quantum processors~\cite{mantha2024pilotquantum}. Kubernetes provides unified resource management and scheduling, allowing dynamic allocation of resources according to workload demands. Declarative pipeline definitions encapsulate task orchestration, dependencies, and execution logic, minimizing manual intervention. Advanced scheduling, auto-scaling, and fault-tolerance mechanisms support large ensembles, parameter sweeps, and data-intensive preprocessing with high throughput and utilization~\cite{10030052}. Integration with workflow engines such as Argo \cite{argo_workflows}, Nextflow \cite{DiTommaso2017nextflow}, and Kubeflow \cite{kubeflow} further enables automation, provenance tracking, and hybrid cloud/HPC interoperability.

These features make Kubernetes a powerful platform for reproducible, scalable, and automated hybrid scientific workflows, providing a bridge between traditional HPC pipelines and modern cloud-native computing frameworks.

\begin{figure*}[t]
    \centering
    \includegraphics[width=0.73\linewidth]{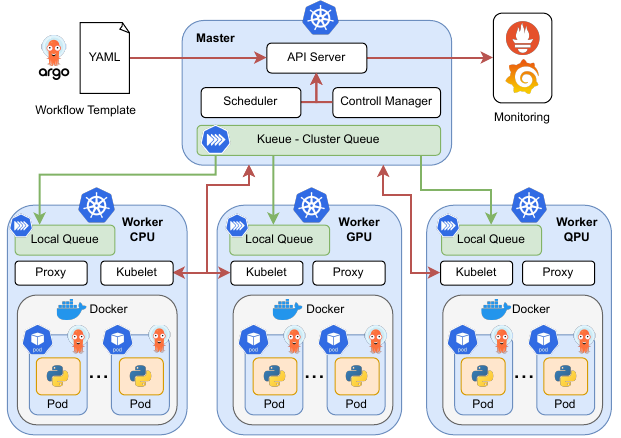}
    \caption{Kubernetes-native hybrid quantum–HPC workflow architecture.}
    \label{fig:schema_k8s}
\end{figure*}

\section{System Architecture}

We present a cloud-native framework for orchestrating hybrid quantum–HPC workflows, shown in Fig.~\ref{fig:schema_k8s}, designed to provide scalable, reproducible, and observable execution across heterogeneous resources. The system unifies CPUs, GPUs, and QPUs under a single orchestration layer, abstracting hardware differences while exposing a uniform interface for users~\cite{burns2016design,argo_workflows}.

Experiments are defined declaratively using YAML-based workflow specifications and submitted to the Kubernetes cluster. Tasks are executed as containerized Pods, scheduled across available compute nodes, with real-time metrics collected via integrated monitoring services, providing visibility into workflow progress, resource utilization, and system performance.

The architecture integrates four key components: (i) \textbf{Kubernetes} as the orchestration backbone, managing compute resources, isolation, and scalability; (ii) \textbf{Argo Workflows}, representing hybrid quantum–classical pipelines as directed acyclic graphs (DAGs)~\cite{argo_workflows}; (iii) \textbf{Kueue}, a workload-aware scheduler that allocates tasks efficiently across CPUs, GPUs, and QPUs~\cite{kueue}; and (iv) \textbf{Prometheus and Grafana} for continuous monitoring and observability~\cite{prometheus,grafana}. This layered design reduces coupling between infrastructure, workflow definition, and scheduling policies: Kubernetes abstracts infrastructure, Argo manages workflow logic, Kueue enforces fair and efficient scheduling, and Prometheus/Grafana provides fine-grained observability.

The system employs a \textbf{master--worker architecture} to enable modularity, and a clear separation of responsibilities~\cite{burns2016kubernetes,argo_workflows,kueue}. The \textit{master node} hosts the Kubernetes control plane, the Argo workflow engine, and the Kueue queue manager, orchestrating task submission, scheduling, and monitoring. \textit{Worker nodes} execute computational workloads.

\subsection{Kubernetes as the Orchestration Backbone}

Kubernetes provides the core orchestration layer of our system, abstracting heterogeneous compute resources and enabling reproducible execution of hybrid quantum--classical workflows~\cite{burns2016design}. Workloads are encapsulated as \texttt{Pods}, which package containers using Docker~\cite{merkel2014docker} and can be scheduled dynamically across cluster nodes, decoupling application logic from specific hardware deployments.

In our proof-of-concept implementation, the Kubernetes cluster comprises a heterogeneous pool of nodes labeled according to their dominant computational role (\texttt{cpu}, \texttt{gpu}, \texttt{qpu}). These labels are leveraged through \texttt{nodeSelector} directives, orchestrated via Argo Workflows, to steer workflow stages toward appropriate execution environments. For example, GPU-intensive classical simulation tasks can be explicitly targeted to GPU-capable nodes as follows:

\begin{center}
\begin{tcolorbox}[colback=black!5!white, colframe=black!75!black,
                  sharp corners, boxrule=0.8pt, left=2mm, right=2mm,
                  top=1mm, bottom=1mm, width=0.25\textwidth,
                  halign=center, valign=center]
\footnotesize
\begin{verbatim}
nodeSelector:
  resource_type: gpu
\end{verbatim}
\end{tcolorbox}
\end{center}

While this approach proved sufficient for demonstrating end-to-end hybrid workflow execution, it is not expected to scale as a long-term or production-grade resource allocation strategy. Explicit node targeting tightly couples workloads to infrastructure details and bypasses Kubernetes' native scheduling capabilities. A more robust solution would rely on resource requests and limits, allowing the scheduler to match workloads to available devices dynamically. This model is already well established for accelerators such as NVIDIA and AMD GPUs, for example:

\begin{center}
\begin{tcolorbox}[colback=black!5!white, colframe=black!75!black,
                  sharp corners, boxrule=0.8pt, left=2mm, right=2mm,
                  top=1mm, bottom=1mm, width=0.25\textwidth,
                  halign=center, valign=center]
\footnotesize
\begin{verbatim}
resources:
  limits:
    nvidia.com/gpu: 1
    amd.com/gpu: 1
\end{verbatim}
\end{tcolorbox}
\end{center}

However, an equivalent standardized mechanism does not yet exist for QPUs, which currently lack first-class support as schedulable resources within Kubernetes.

Looking forward, future iterations of this architecture will leverage Kubernetes Dynamic Resource Allocation (DRA), which became generally available in Kubernetes~1.34. DRA introduces a more expressive resource model based on device \emph{classes} and workload \emph{claims}, allowing system administrators to describe heterogeneous resources---such as different CPU architectures, GPU models, or QPUs with varying quality-of-service (QoS) characteristics---while leaving the matchmaking, allocation, and device setup to the scheduler and device plugins~\cite{kubernetesDRA}. Under this model, workloads request access to abstract resource classes rather than targeting specific nodes, enabling late binding, improved portability, and more efficient cluster utilization.


\subsection{Integration of Quantum Resources in Kubernetes}

Integrating QPUs into Kubernetes-based hybrid workflows can be achieved via \textbf{on-premise} or \textbf{cloud-based} configurations~\cite{Preskill2018quantumcomputingin,McClean_2016,mantha2024pilotquantum}.

In the on-premise configuration, a Kubernetes worker node corresponds to the quantum computer’s access node. Workloads scheduled on this node are executed directly on the local QPU, providing low-latency interaction and tight coupling between classical and quantum resources.

For cloud-based QPU access, Kubernetes nodes act as clients interfacing with the quantum provider’s API. Authentication credentials, such as API tokens, are securely managed using Kubernetes \textit{Secrets} and mounted as read-only volumes. For example:

\begin{lstlisting}[language=yaml, caption={Workflow metadata and mounted secrets for cloud-based QPU access.}]
env:
  - name: IQM_TOKENS_FILE
    value: "/mnt/shared/iqm/tokens.json"
volumeMounts:
  - name: shared-data
    mountPath: /mnt/shared/
  - name: iqm-tokens
    mountPath: /mnt/shared/iqm
    readOnly: true
\end{lstlisting}

This approach ensures secure authentication while enabling containerized tasks to submit jobs, monitor execution, and retrieve results from cloud QPUs. By supporting both local and remote configurations, Kubernetes-based orchestration supports both types of quantum resources allowing a unified, reproducible, and interoperable framework for managing quantum resources.

\subsection{Workflow Management with Argo Workflows}

\begin{figure}[h]
    \centering
    \includegraphics[width=0.6\linewidth]{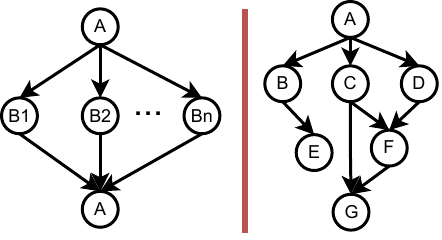}
    \caption{Example of an Argo DAG representation: the left shows a generic Fan Out/Fan In workflow, while the right panel illustrates a specific application workflow.}
    \label{fig:argo}
\end{figure}

Kubernetes provides the orchestration backbone, while \textit{Argo Workflows} adds a declarative layer for defining hybrid quantum--classical pipelines~\cite{argo_workflows}. Workflows are specified in YAML and represented as Directed Acyclic Graphs (DAGs), where each node encapsulates a containerized task. See Fig \ref{fig:argo} for examples. This model captures modularity, data dependencies, and stage boundaries inherent in hybrid HPC and quantum workflows.

The DAG-based design enables reproducible, composable, and transparent experiments. Workflows can be version-controlled and re-executed with identical semantics, templates can be reused or extended for new pipelines, and each stage is explicitly defined for fine-grained traceability. Containerized execution combined with declarative specification ensures scalable and automated orchestration of heterogeneous workloads within Kubernetes, facilitating end-to-end reproducibility in hybrid quantum-HPC research.

\subsection{Queue-Based Scheduling with Kueue}

While Kubernetes provides a default scheduler, its policies are not optimized for heterogeneous hybrid environments with scarce resources~\cite{kueue}. To address this, \textit{Kueue} introduces queue-based, workload-aware scheduling that dynamically manages CPU, GPU, and QPU tasks according to resource availability and priority. Workload pieces are assigned to queues, ensuring fairness and responsiveness across heterogeneous nodes while respecting workflow dependencies.

\begin{lstlisting}[language=yaml, caption={Kueue ResourceFlavor definition for GPU.}, label={lst:kueue-resourceflavor}]
apiVersion: kueue.x-k8s.io/v1beta1
kind: ResourceFlavor
metadata:
  name: gpu-flavor
spec:
  nodeSelector:
    resource_type: gpu
\end{lstlisting}

\begin{lstlisting}[language=yaml, caption={Kueue LocalQueue definition for GPU jobs.}, label={lst:kueue-localqueue}]
apiVersion: kueue.x-k8s.io/v1beta1
kind: LocalQueue
metadata:
  namespace: quantum-workflows
  name: queue-gpu
spec:
  clusterQueue: cluster-queue 
\end{lstlisting}

\begin{lstlisting}[language=yaml, caption={Submission referencing the GPU LocalQueue.}, label={lst:kueue-submission}]
...
  - name: execute-subcircuits-gpu
    metadata:
      labels:
        kueue.x-k8s.io/queue-name: queue-gpu
...
\end{lstlisting}

The listings below illustrate the key components of a Kueue-based scheduling setup. Listing~\ref{lst:kueue-resourceflavor} defines a \texttt{ResourceFlavor} for GPU nodes, specifying a node selector to identify nodes equipped with GPU resources. Listing~\ref{lst:kueue-localqueue} shows a \texttt{LocalQueue} bound to the GPU resource flavor, which serves as a queue for the GPU worker. Finally, Listing~\ref{lst:kueue-submission} demonstrates how a workflow step is submitted to the GPU queue via a label reference, enabling Kueue to schedule the task according to resource availability and queue policies.

\subsection{Monitoring and Observability}

Hybrid quantum--classical workflows require comprehensive monitoring to ensure reproducibility, performance, and scalability~\cite{prometheus,grafana}. This system employs a Prometheus--Grafana stack, where Prometheus collects real-time metrics from Kubernetes pods, nodes, and orchestration components, storing them in a time-series database, while Grafana provides interactive dashboards for visualization.

Key metrics include task states (pending, active, succeeded, failed), CPU and GPU utilization, QPU latency, and workflow throughput. Continuous monitoring enables detection of performance bottlenecks, tuning of scheduling parameters, and assessment of scalability across heterogeneous resources. The observability layer thus provides a transparent, operational view of hybrid workflows, supporting reproducible and efficient execution of quantum--classical pipelines.

\section{Proof of Concept: Hybrid Circuit-Cutting Workflow}

Circuit cutting was chosen as a representative hybrid quantum–classical workload to validate the proposed orchestration system~\cite{tejedor2025distributed,harada2023wirecut}. In essence, circuit cutting is a technique that partitions a large $n$-qubit quantum circuit into smaller, overlapping subcircuits (or “fragments”) that can be executed independently on available hardware. These fragments include additional operations that account for the interactions between the cut points, ensuring that when the results are recombined, the expectation values of global observables are preserved.

\begin{figure}[h]
    \centering
    \includegraphics[width=1\linewidth, trim=0 130 0 130, clip]{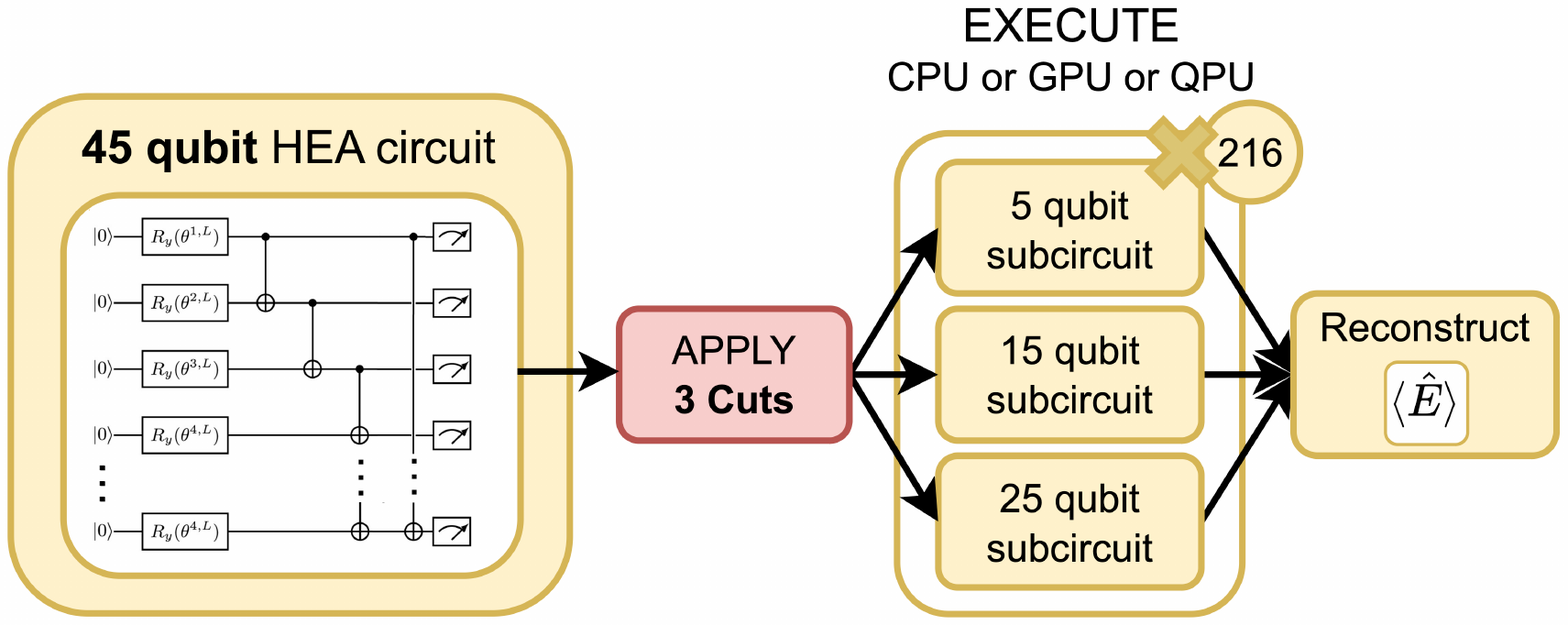}
    \caption{Schematic overview applying 3 cuts in a HEA circuit}
    \label{fig:workflow_overview}
\end{figure}

As illustrated in Figure \ref{fig:workflow_overview}, the workflow executes these subcircuits across CPU, GPU, or QPU backends, depending on the selected resource and circuit size. After execution, a classical post-processing step recombines the fragment outputs to reconstruct the results of the original circuit. This approach allows exploration of quantum workloads that exceed the capacity of individual quantum devices, while also providing a flexible testbed to evaluate orchestration efficiency, workload balancing, and reproducibility across heterogeneous resources.

\subsection{Workflow Pipeline}

The pipeline consists of three containerized stages forming an Argo DAG: \textbf{subcircuit generation}, \textbf{parallel execution}, and \textbf{result reconstruction}. CPU nodes generate subcircuits via a gate-cutting algorithm that identifies optimal cuts and produces independent fragments; this is done using Qdislib \cite{tejedor2025distributed} and Qiskit \cite{qiskit2019}. These fragments are serialized as Python \texttt{pickle} objects and stored in a shared volume for distributed evaluation.

Subcircuits are then distributed across CPU, GPU, and QPU nodes according to a backend-selection policy. For demonstration purposes, we use a simple toy policy that assigns small circuits to a QPU, medium circuits to a CPU, and large circuits to a GPU:

\vspace{5mm}

\begin{lstlisting}[language=Python, caption={Toy backend selection policy for demonstration purposes.}]
def select_backend(num_qubits):
    if num_qubits <= 5:
        return "QPU" # small circuits
    elif num_qubits <= 20:
        return "CPU" # medium circuits
    else:
        return "GPU" # big circuits
\end{lstlisting}

This policy is intentionally simplistic and serves only to illustrate the orchestration workflow. In a realistic setting, assigning large circuits to GPUs while reserving QPUs for only very small qubit sizes would defeat the primary advantage of using quantum hardware. Proper QPU scheduling would instead consider factors such as circuit depth, qubit connectivity, and the potential quantum speedup.

\begin{figure*}[htbp]
  \centering
  \includegraphics[width=0.6\textwidth, trim=50 0 50 0, clip, angle=90]{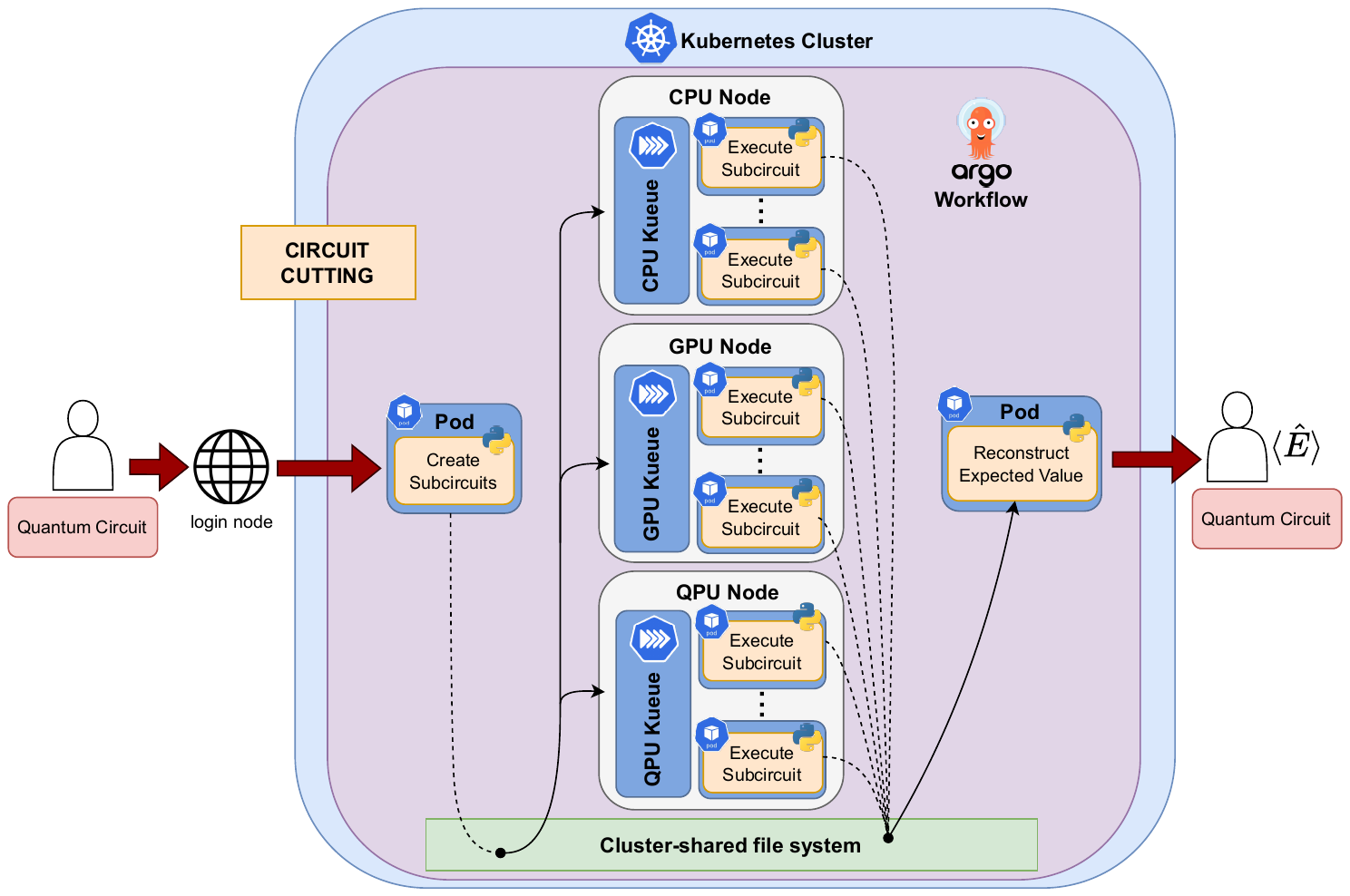}
  \caption{Concurrent subcircuit execution across heterogeneous nodes. Argo manages data dependencies while Kueue balances workloads between CPU, GPU, and QPU resources.}
  \label{fig:parallel_execution}
\end{figure*}

Each subcircuit is executed as a containerized Argo step, with Kueue scheduling tasks based on resource availability and queue priorities, as illustrated in Fig.~\ref{fig:parallel_execution}. Concurrent execution enables overlap of independent tasks, potentially reducing wall-clock time.

CPU nodes then aggregate all subcircuit outputs to reconstruct the global expectation value using the gate-cutting formula, a step that follows the standard circuit-cutting formalism, which is theoretically unbiased, allowing validation against partial or full QPU executions.

\subsection{Argo DAG Definition and Execution}

The hybrid quantum--classical pipeline is defined declaratively in Argo Workflows. Shared volumes provide persistent data storage and secret management, as shown in Listing~\ref{lst:workflow-volumes}.

{\scriptsize
\begin{lstlisting}[language=yaml, caption={Workflow metadata and shared volumes.}, label={lst:workflow-volumes}]
apiVersion: argoproj.io/v1alpha1
kind: Workflow
metadata:
  name: quantum-subcircuits
spec:
  entrypoint: main
  volumes:
    - name: shared-data
      persistentVolumeClaim:
        claimName: shared-pvc
    - name: iqm-tokens
      secret:
        secretName: iqm-tokens
\end{lstlisting}
}

Listing~\ref{lst:workflow-steps} defines the Argo Workflow DAG for the circuit-cutting pipeline. After generating all subcircuits, the workflow launches parallel CPU, GPU, and QPU execution branches, each instantiating 216 parameterized tasks via \texttt{withSequence}. Kubernetes and Kueue manage concurrency and resource placement across heterogeneous backends. The final \texttt{reconstruct} step synchronizes all executions and performs the classical reconstruction.

From a user perspective, the workflow is launched with a single command:

\begin{lstlisting}[language=bash]
kubectl apply -f circuit_cutting_workflow.yaml
\end{lstlisting}

This setup enables parallel execution of hybrid quantum–classical tasks while maintaining reproducibility and easy deployment within a Kubernetes cluster.

\begin{lstlisting}[language=yaml, caption={Main workflow steps.}, label={lst:workflow-steps}]
templates:
  - name: main
    steps:
      - - name: create-subcircuits
          template: create-subcircuits
      - - name: execute-subcircuits-cpu
          template: execute-subcircuits-cpu
          arguments:
            parameters:
              - name: index
                value: "{{item}}"
          withSequence: {count: 216}
        - name: execute-subcircuits-gpu
          template: execute-subcircuits-gpu
          arguments:
            parameters:
              - name: index
                value: "{{item}}"
          withSequence: {count: 216}
        - name: execute-subcircuits-qpu
          template: execute-subcircuits-qpu
          arguments:
            parameters:
              - name: index
                value: "{{item}}"
          withSequence: {count: 216}
      - - name: reconstruct
          template: reconstruct
\end{lstlisting}

\section{Discussion}\label{sec:discussion}

The proof-of-concept deployment of our Kubernetes-native hybrid quantum--HPC workflow demonstrates the feasibility and advantages of using cloud-native orchestration to coordinate classical and quantum resources. 
The workflow assigns distinct roles to CPUs, GPUs, and QPUs, aligning each stage with the most suitable execution backend. 
Subcircuit generation and reconstruction are handled centrally on CPU nodes, simplifying data management and synchronization.
GPU-backed execution enables state-vector simulations that would be impractical on CPU-only resources, and QPU endpoints integrated seamlessly with remote hardware. Argo Workflows and Kueue enabled concurrent execution of independent subcircuits, exploiting both intra- and inter-stage parallelism. Monitoring metrics indicate sustained CPU and GPU utilization during execution, without prolonged idle periods and the queue-based scheduler ensures ensures fair allocation of resources through Kueue’s queue-based admission control and quota mechanisms while avoiding conflicts between classical and quantum workloads.

Reproducibility and transparency were achieved through containerization, Argo DAGs, and YAML-based workflow specifications, which provide version-controlled, declarative definitions for experiments. Persistent volumes and Kubernetes Secrets enabled consistent data exchange and secure handling of QPU credentials. Fine-grained instrumentation using Prometheus and Grafana allowed identification of bottlenecks, such as GPU concurrency limitations and I/O saturation during reconstruction, establishing a fully observable foundation for hybrid workloads.

The workflow successfully integrated both on-premise and cloud QPU backends, demonstrating that Kubernetes can abstract quantum resources in a manner analogous to classical accelerators, thereby simplifying the development of hybrid quantum--classical workflows. Despite these advantages, several limitations were observed. In particular, the lack of fine-grained GPU sharing constrained the degree of concurrent simulations, as GPUs were allocated as exclusive devices rather than partitioned resources. This limitation is not fundamental to the Kubernetes model and is being actively addressed through the Dynamic Resource Allocation (DRA) framework discussed in Sec.~IV-B. DRA allows system administrators to expose not only full devices (e.g., GPUs or QPUs), but also resource classes corresponding to fractional or virtualized portions of a device. This capability opens the door to more flexible scheduling strategies, where multiple workloads may share a single accelerator according to well-defined quality-of-service constraints.

More broadly, DRA enables a shift toward more expressive and hardware-agnostic resource specifications. In future deployments, workloads may request resources in terms of abstract capabilities---such as numbers of CUDA cores, GPU memory slices, or quantum-specific metrics like shot budgets or execution windows rather than binding to concrete devices. Such abstractions are particularly well aligned with the heterogeneous and evolving nature of quantum hardware, and they offer a path toward more efficient utilization of both classical and quantum accelerators.

Additional limitations include latency, I/O bottlenecks arising from persistent volume usage at larger scales, and the absence of predictive scheduling or automated backend selection based on real-time telemetry. While Kueue provides fair queue-based scheduling, integrating performance-aware or telemetry-driven decision-making remains an open challenge. Addressing these limitations constitutes a natural direction for future work and is essential for scaling hybrid quantum--classical workflows toward production-grade deployments.

\section{Conclusions}\label{sec:conclusions}

This work presents a scalable, reproducible, and observable framework for orchestrating hybrid quantum–classical workflows using Kubernetes, Argo, and Kueue. The proof-of-concept demonstrates seamless orchestration of heterogeneous resources and effective execution of distributed quantum circuits, providing a foundation for a wide range of hybrid algorithms, including circuit cutting, as proved, sample-based diagonalization \cite{sample-based-diagonalization}, and other approaches that combine quantum with classical computation. In addition, the framework naturally supports workloads arising in Quantum Machine Learning (QML), which represent a distinct class of hybrid quantum–classical problems requiring iterative quantum evaluations and classical optimization. By demonstrating effective orchestration of distributed quantum workloads, this platform enables flexible and efficient experimentation with complex hybrid quantum–classical strategies across multiple application domains.

While Slurm remains the de facto standard for batch scheduling in HPC and is widely adopted by quantum computing providers, it is primarily designed for static, job-level execution. Kubernetes, in contrast, enables fine-grained orchestration of multi-stage, containerized workflows with dynamic resource allocation, integrated monitoring, and native support for heterogeneous services. A detailed comparison between Slurm and Kubernetes is beyond the scope of this work, as such an evaluation would require a broader architectural and operational analysis than intended here. Moreover, the two systems are not mutually exclusive: Kubernetes can be integrated alongside existing HPC schedulers, enabling hybrid deployments in which batch-managed compute resources coexist with cloud-native orchestration layers.
The contribution of this work is therefore not to position one system as a replacement for the other, but to demonstrate that cloud-native orchestration enables execution patterns such as dynamic hybrid pipelines, service-level integration of quantum processing units, and workflow-level observability.

Future developments will focus on several directions: enhancing scheduling intelligence through predictive or performance-aware algorithms, supporting parameterized variational circuits and iterative quantum-classical loops, reducing I/O overhead via in-memory or distributed storage solutions, and extending orchestration across multiple clusters and HPC infrastructures. These improvements will strengthen the flexibility, efficiency, and scientific utility of cloud-native hybrid quantum-HPC systems, paving the way for broader adoption of hybrid workflows in research and production environments.

\section*{Acknowledgements}
We gratefully acknowledge the CERN Openlab program for supporting M.T.’s internship. M.G., C.T., is supported by CERN through the CERN Quantum Technology Initiative. We also thank Diana Gaponcic and Jack Charlie Munday for their valuable help.

\bibliographystyle{IEEEtran}
\bibliography{main}

\end{document}